\documentclass[epj]{webofc}
\usepackage[varg]{txfonts}   
\usepackage{graphicx,color} 
\usepackage{bm}       
\usepackage{amsmath}  
\usepackage{amssymb}
\usepackage{float}
\woctitle{21st International Conference on Few-Body Problems in Physics}
\begin{document}
\title{Scaling functions of two-neutron separation energies of $^{20}C$ with finite range potentials}
\author{M. A. Shalchi\inst{1}\fnsep\thanks{\email{shalchi@ift.unesp.br}} \and
        M. R. Hadizadeh\inst{2}\fnsep\thanks{\email{hadizadm@ohio.edu}} \and
        M. T. Yamashita\inst{1}\fnsep\thanks{\email{yamashita@ift.unesp.br}} \and
        Lauro Tomio\inst{1,3}\fnsep\thanks{\email{tomio@ift.unesp.br}} \and
        T. Frederico\inst{4}\fnsep\thanks{\email{tobias@ita.br}}                 
}

\institute{Instituto de F\'{\i}sica Te\'orica, UNESP, 01140-070, S\~ao Paulo, SP, Brazil
\and
           Institute for Nuclear and Particle Physics and Department of Physics and Astronomy, 
Ohio University, Athens, OH, USA 
\and
           Centro de Ci\^encias Naturais e Humanas, Universidade Federal do ABC, 09210-580,
Santo Andr\'e, SP, Brazil
\and
           Instituto Tecnol\'ogico de Aeron\'autica, DCTA, 12228-900, S\~ao
Jos\'e dos Campos, SP, Brazil.
          }

\abstract{
The behaviour of an Efimov excited state is studied within a three-body Faddeev formalism for 
a general neutron-neutron-core system, where neutron-core is bound and neutron-neutron is 
unbound, by considering zero-ranged as well as finite-ranged two-body interactions. 
For the finite-ranged interactions we have considered a one-term separable Yamaguchi potential.
The main objective is to study range corrections in a scaling approach, with focus in the exotic 
carbon halo nucleus $^{20}C$.
}
\maketitle
\section{Introduction and Model}\label{intro}
In the present contribution our aim is to consider a general scaling model for three-body system 
with two non-identical ones near the unitary limit (when one or both two-body scattering lengths
have absolute values very large), by considering bound neutron-core (nc) subsystem.
The approach is exemplified for the exotic carbon halo nucleus $^{20}C$, where an 
Efimov \cite{Efimov} excited energy state is studied within a three-body Faddeev formalism for 
$n-n-^{18}C$ (also named neutron-neutron-core, $nnc$, for more general neutron-rich exotic 
nuclei). More details on related studies, one can obtain from \cite{AmorimPRC97,Marcelo,Hammer,Tobias,Phillips1,Phillips2} 
and references therein. 
Therefore, the first task was to solve the coupled $s-$wave 
Faddeev integral equation using a zero-range potential to obtain the two-neutron ($nn$) 
separation energies for the ground and excited states of $^{20}C$. Next, by fixing the energy 
of the virtual state of the $nn$ subsystem to $E_{nn}=-143$ keV, we vary a three-body 
parameter in order to obtain different $^{19}C$ binding energies for the $n-^{18}C$ subsystem, 
such that we can also reproduce the $^{20}C$ ground-state energy, namely $E^{(0)}_{^{20}C}$.
This procedure is followed by the calculation of the first excited-state energy $E^{(1)}_{^{20}C}$. 
In this way, we are able to obtain a scaling plot, which is conveniently given in dimensionless 
quantities, as 
$\displaystyle \sqrt{\left[
{E^{(1)}_{^{20}C}-E_{^{19}C}}\right]/{E^{(0)}_{^{20}C}}}$ versus 
$\displaystyle\sqrt{{E_{^{19}C}}/{E^{(0)}_{^{20}C}}}$ \cite{Tobias,scale}.
In order to study the range corrections, the same above steps are repeated, with the zero-range 
interaction being replaced by a separable Yamaguchi potential\cite{Yamaguchi}, with an appropriate range 
parameter $\beta$.

In the following we present the corresponding model-formalism, where for the two-body interactions
we have used zero-ranged potentials (constant in momentum space), renormalized to give us the
corresponding energies of the two-body subsystems ($nn$ and $nc$), as well as finite-range
interactions, given by one-term separable Yamaguchi form-factors.  In the case of the finite-ranged
interactions, the two-body potential, with strength $\lambda$ and range-parameter $\beta$, is given by
\begin{equation}
 v(p,p')=\lambda \, g(p) \, g(p') =\lambda \,\left(\dfrac{1}{p^2+\beta^2} \right) \left(\dfrac{1}{p'^2+\beta^2} \right).
 \label{v-sep}\end{equation}
The corresponding two-body $t-$matrix is also given analytically in a separable form, as follows:
\begin{equation}
t( p,p',E)=\tau(E)\,g(p) \, g(p'),
\label{t-sep}\end{equation}
where
\begin{equation}
\tau(E)=\left [ \lambda^{-1}-\int d^3p'' \, g^2(p'') \, G_0(p'') \right]^{-1}=\left (\lambda^{-1}+\dfrac{2\pi^2\mu}{\beta(\beta\pm k)^2} \right )^{-1}  .
\end{equation}
The plus and minus signs are for bound and virtual states, respectively.  
$\mu$ is reduced mass and we are considering units such that $\hbar c=1$.
The parameters of the potential $ \lambda$ and $\beta$ are given by the corresponding values of the subsystem energies,
since the two-body $t-$matrix has poles for bound and virtual state energies. Next, the above definitions will be identified with
 the corresponding two-body subsystems by the respective labels $nn$ and $nc$.

In our treatment for the two-neutron halo nuclei, we consider the two neutrons with masses given by $m_n\equiv m$ and
the core with mass $m_A\equiv A \,m$.  
Such system can be studied in momentum space by the Faddeev formalism as a three-body problem with two different mass particles. 
With the assumption that the third particle is the core, with particles 1 and 2 being the neutrons, the coupled Faddeev equations for
the $|\psi\rangle^{nn}$ and $|\psi\rangle^{nc}$ components of the subsystems $nn$ and $nc$ are given by 
\begin{eqnarray} \label{eqa1}
|\psi\rangle^{nn}&= &G_{0nn} \,t_{nn} \,  \left (|\psi\rangle^{cn}+|\psi\rangle^{nc} \right)  = 2 \, G_{0nn} \,t_{nn} \, |\psi\rangle^{nc},  \cr
|\psi\rangle^{nc} &=& G_{0nc} \,t_{nc} \, (|\psi\rangle^{nn}+|\psi\rangle^{nc}) 
=G_{0nc} \,t_{nc} \, (|\psi\rangle^{nn}+P_{12}|\psi\rangle^{nc}),
\end{eqnarray}
where $G_{0nn}$ ($G_{0nc}$) and $t_{nn}$ ($t_{nc}$) are the three-body free propagator and two-body $t-$matrix for the
$nn$ ($nc$) subsystem, respectively. After partial-wave projection, in momentum space, for the $s-$wave, we have the following 
definitions for the wave-function components,
\begin{eqnarray} \label{spectator_coupled}
 \psi_{nn}(p,q) &=&4\pi \, G_{0nn} \,(p,q) \, g_{nn}(p) \,F_1(q) , \cr
 \psi_{nc}(p,q)&=& 4\pi \, G_{0nc} \,(p,q) \, g_{nc}(p) \, F_2(q).
\end{eqnarray}
with the corresponding Faddeev spectator functions, $F_1(q)$ and $F_2(q)$,
given by {\small
\begin{eqnarray}
\label{eqmain}
 F_1(q)&=& 4\pi \, \tau_{nn} \left (E-\frac{q^2(A+2)}{4mA} \right )\int dq' \, q'^2 \int dx \,\frac{2m}{2m E-{(1+1/A)}q^2-{2q'^2}-{2qq'x}}F_2(q'), \cr
 F_2(q)&=&2\pi \, \tau_{nc} \left(E-\frac{q^2(A+2)}{2m(A+1)} \right )\int dq' \, q'^2 \int dx \,\frac{2m}{2mE-{2q^2}+{(1+1/A)}q'^2+{2qq'x}}F_1(q')\\
&+&2\pi  \, \tau_{nc} \left (E-\frac{q^2(A+2)}{2m(A+1)}\right)\int dq' \, q'^2 \int dx \,\frac{2mA}{2mA E-{{(A+1)}(q^2-q'^2)+{2qq'x}}}F_2(q')  . \nonumber 
\end{eqnarray} }
One can easily verify, from Eqs.~(\ref{v-sep}) and (\ref{t-sep}), that the results given by above formalism for separable one-term Yamaguchi interaction 
must converge to the zero-range results for very large values of $\beta$.
In agreement with this expectation, next, we present some sample results for two cases where the $nn$ virtual-state energy is fixed.

\section{Results - Scaling plots and range corrections}
 
 \subsubsection*{(a) Three-body scaling plot for $E_{nn}=0$}
 
\begin{figure}[H]
\centering
\sidecaption
\includegraphics[width=6cm,clip]{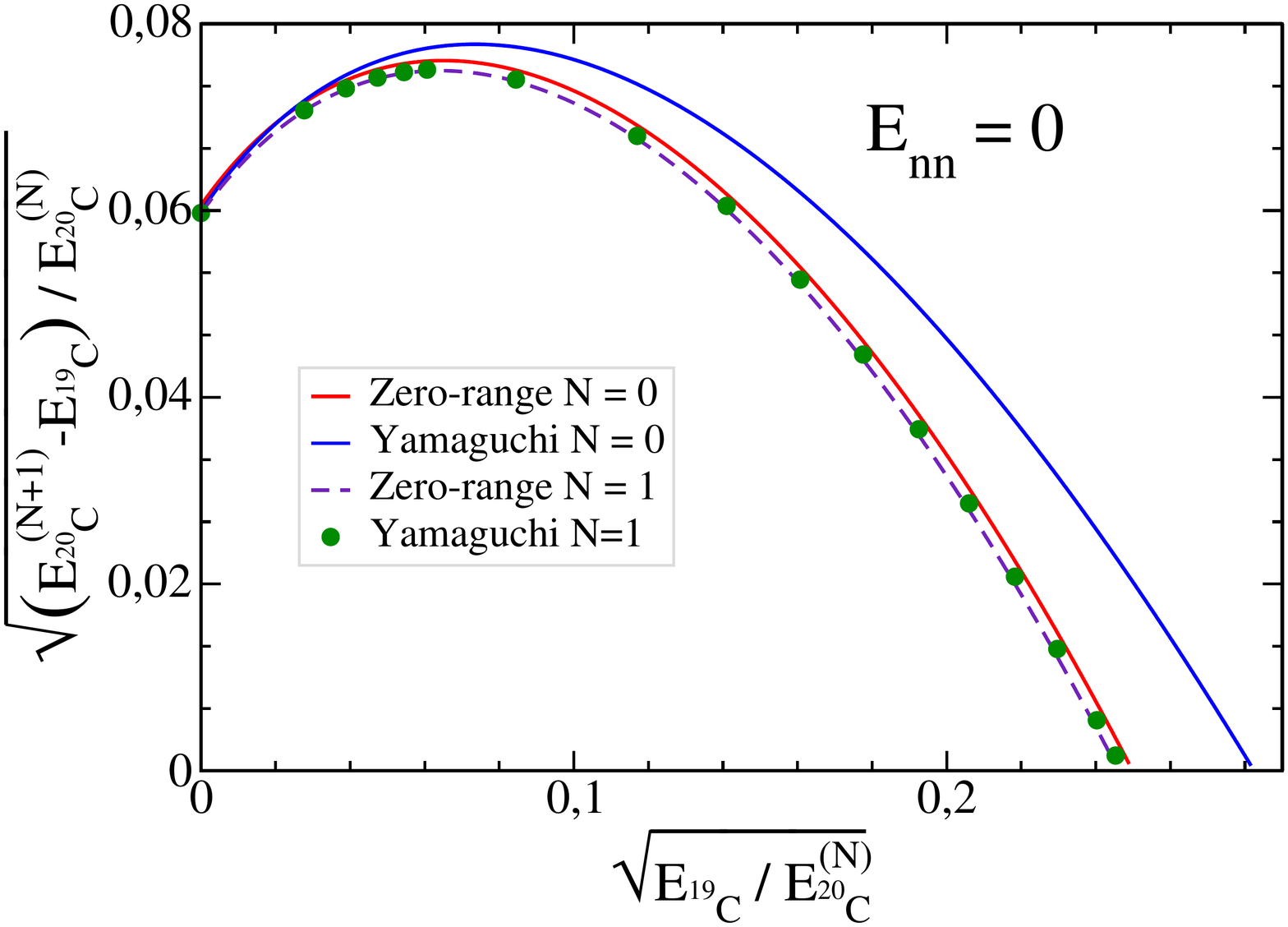}
\caption{Scaling plots for the first and the second cycles, with $E_{nn}=0$. The range effect in the scaling plot given by the Yamaguchi 
potential disappears in the second cycle, i.e. $N=1$.
}
\label{fig-1}       
\end{figure}

 In Fig.~\ref{fig-1} we have presented the scaling plots, which are correlating the binding energies of two successive $^{20}C$ states 
 $E_{^{20}C}^{(N)}$ and $E_{^{20}C}^{(N+1)}$, for the case that the virtual-state energy of the $nn$ subsystem is zero ($E_{nn}=0$). 
In this case, $N=0$ corresponds to the ground-state. The scaling plots are shown for zero-range and Yamaguchi potentials and for two cycles,
with $N=0$ and $N=1$, implying that we need to obtain the first two excited states ($N=1,2$). 
As we can see, the first cycle, for zero-range and Yamaguchi potentials, have some deviation, which are showing clearly the range effect 
obtained by using the Yamaguchi finite-range interaction.
However, for the second cycle ($N=1$), the scaling plots for zero-range and Yamaguchi potentials are almost the same, which is not surprising 
since the ratio between the range and scattering length is very small and consequently the Yamaguchi scaling plot approaches the results obtained 
with the zero-range interaction.
 
\subsubsection*{(b) Three-body scaling plot for $E_{nn}=-143 \, \text{keV}$}

\begin{figure}[H]
\centering
\sidecaption
\includegraphics[width=6cm,clip]{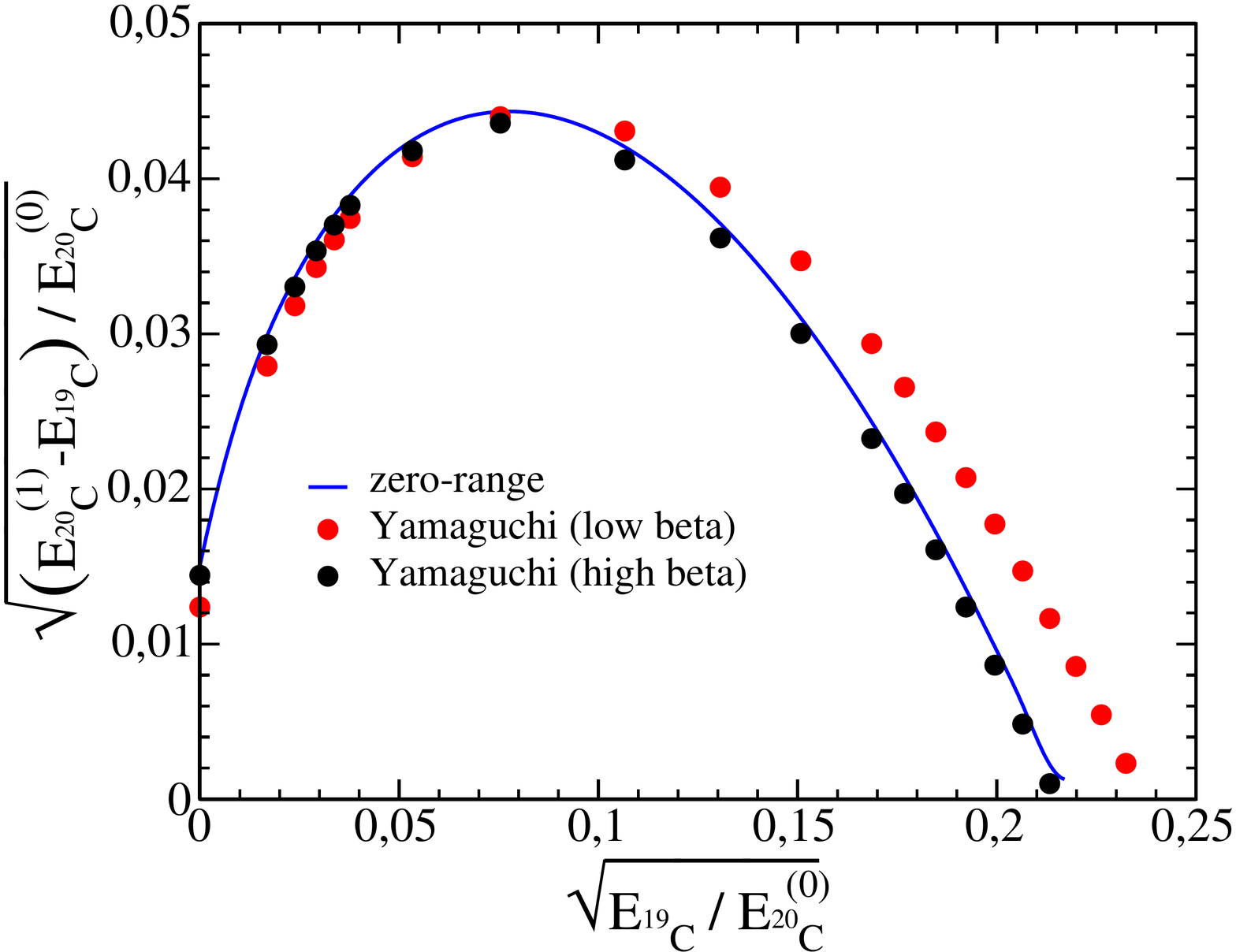}
\caption{Scaling plots for zero-range and Yamaguchi potentials, with $E_{nn}=-143$ keV. The scaling plot for the Yamaguchi potential goes to zero-range 
results for large values of $\beta$ (lower values of the effective range).}
\label{fig-2}       
\end{figure}

In this section we have presented our results for realistic case, when $E_{nn}=-143$ keV. As it can be verified from our formulation, for large values of parameter $\beta$, the coupled Faddeev equations for Yamaguchi potential goes to zero-range equations. As we can see in Fig.~\ref{fig-2}, the scaling plot for Yamaguchi potential for high values of $\beta$, or low values of range, is almost the same with zero-range results.
This is very useful test to numerically verify the validity of our calculations.

\subsubsection*{(c) Effective range correction}
\begin{figure}[H]
\centering
\sidecaption
\includegraphics[width=5.5cm,clip]{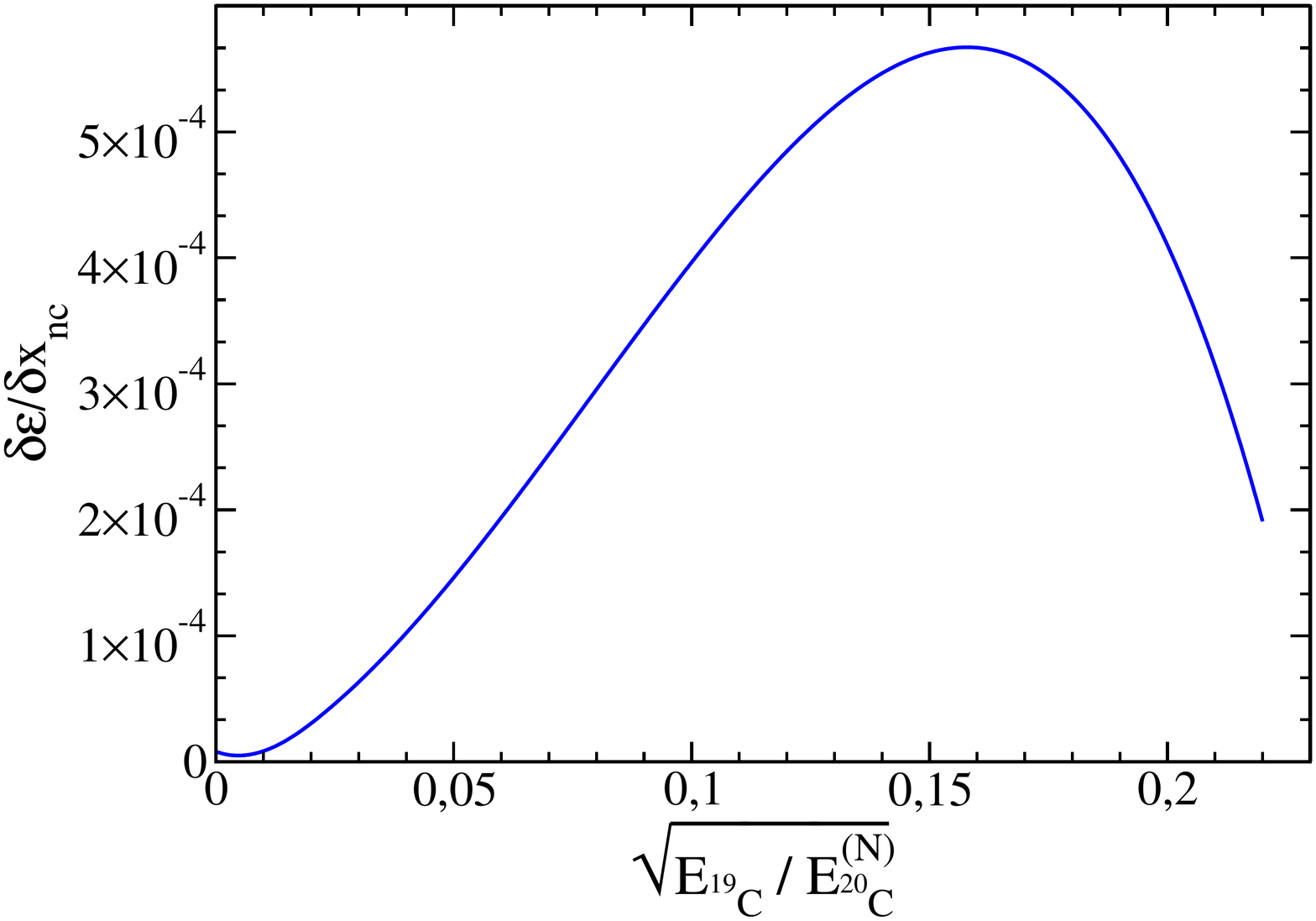}
\caption{The derivation of $^{20}C$ scaling plot, given in Eq. (\ref{expansion}), for $E_{nn}=$-143 keV as a function of the $nc$ subsystem energy.}
\label{fig-3}       
\end{figure}

The correlation between the binding energy of two successive three-body states is given by a scaling function varying with the energies and  
ranges of the subsystems. 
By choosing a very small range for the $nn$ subsystem, one can expand the scaling function in terms of the range of the $nc$ subsystem, 
i. e. $x_{nc}=r_{nc}\sqrt{2\mu_{nc}E_{nnc}^{(N)}}$. For leading order expansion one can have
\begin{eqnarray}\label{expansion}
 \dfrac{E_{nnc}^{(N+1)}}{E_{nnc}^{(N)}}&=&\epsilon \left (-\sqrt{\dfrac{E_{nn}}{E_{nnc}^{(N)}}},\sqrt{\dfrac{E_{nc}}{E_{nnc}^{(N)}}},
 r_{nn}\sqrt{2\,\mu_{nn}\,E_{nnc}^{(N)}},r_{nc}\sqrt{2\,\mu_{nc}\,E_{nnc}^{(N)}} \right ) \cr
 &=&\epsilon \left (-\sqrt{\dfrac{E_{nn}}{E_{nnc}^{(N)}}},\sqrt{\dfrac{E_{nc}}{E_{nnc}^{(N)}}} \right )
 +\dfrac{\delta \epsilon}{\delta x_{nc}} r_{nc}\sqrt{2\,\mu_{nc}\,E_{nnc}^{(N)}} 
\end{eqnarray}
In Fig~\ref{fig-3} we show the results for the derivative $\dfrac{\delta \epsilon}{\delta x_{nc}} r_{nc}$ of this scaling function in terms of the energy of the $nc$ subsystem. Our preliminary results were computed using large values of $\beta$ for $nn$ interaction, while  for the $nc$ potential we use for low $\beta$ values. We compute the range effect in Fig~\ref{fig-3}, which comes essentially from
the $nc$ interaction, as the large values of $\beta$ for the $nn$ potential gives small effective range. Our calculation is done for the Yamaguchi potential, and should be viewed as a preliminary calculation, indicating the range correction pattern in the scaling plot of  
Fig.~\ref{fig-2}.  We have only looked to the range effect in the $nc$ potential, and with a particular form factor for the separable potential.
We intend to study different form factors, in order to implement the range correction directly in the set of renormalised subtracted zero-range 
integral equation. This will allow us to obtain a model independent analysis of the range corrections in the energy scaling plots, namely the 
correlation between the energy of two close states. Other properties as sizes and momentum distribution should also be investigated in the future.
\begin{acknowledgement}
\vspace{-0.2cm}We acknowledge partial financial support from the Brazilian agencies FAPESP, CNPq
and CAPES.
\vspace{-0.2cm}
\end{acknowledgement}
%
%
%

\end{document}